

\input harvmac

\def\s{\sigma}
\def\L{\Lambda}

\def\d{{\rm d}}
\def\IR{\relax{\rm I\kern-.18em R}}
\font\cmss=cmss10 \font\cmsss=cmss10 at 7pt
\def\IZ{\relax\ifmmode\mathchoice
{\hbox{\cmss Z\kern-.4em Z}}{\hbox{\cmss Z\kern-.4em Z}}
{\lower.9pt\hbox{\cmsss Z\kern-.4em Z}}
{\lower1.2pt\hbox{\cmsss Z\kern-.4em Z}}\else{\cmss Z\kern-.4em Z}\fi}

\def\delbar{\overline\del}
\def\inbar{\vrule height1.5ex width.4pt depth0pt}
\def\IC{\relax\thinspace\hbox{$\inbar\kern-.3em{\rm C}$}}

\def\-{\hphantom{-}}
\def\pd#1#2{{{\partial #1}\over{\partial #2}}}

\noblackbox

\Title{\vbox{\baselineskip12pt
\hbox{NEIP92--004}
\hbox{hepth@xxx/9210021}}}
{\vbox{\vskip-0.4in\centerline{Duality Symmetries from}
       \vskip2pt
       \centerline{Non--Abelian Isometries in}
       \vskip2pt
       \centerline{String Theory$^*$}}}%
\footnote{}{$^*$Supported by
     the Swiss National Foundation. E--mail: delaossa and
           quevedo@iph.unine.ch.}

{\vbox{\centerline{Xenia C. de la Ossa}
       \vskip2pt
       \centerline{and}
       \vskip2pt
       \centerline{Fernando Quevedo}}}

\bigskip
{\vbox{\baselineskip12pt
\centerline{Institut de Physique}
\centerline{Universit\'e de Neuch\^atel}
\centerline{CH-2000 Neuch\^atel}
\centerline{Switzerland.}}}

\vskip .4in
\vbox{\baselineskip14pt
\noindent In string theory it is known that abelian isometries
in the $\s$--model lead to target space duality. We generalize
this duality to backgrounds with {\it non--abelian} isometries.
The procedure we follow consists of gauging the isometries of
the original action and constraining the field strength $F$ to
vanish.  This new action generates dual theories by integrating
over either the Lagrange multipliers that set $F=0$ or the gauge
fields.  We find that this new duality transformation maps
spaces with non--abelian isometries to spaces that may have no
isometries at all.  This suggests that duality symmetries in
string theories need to be understood in a more general context
without regard to the existence of continuous isometries on the
target space (this is also indicated by the existence of duality
in string compactifications on Calabi--Yau manifolds which have
no continuous isometries).  Physically interesting examples to
which our formalism apply are the Schwarzschild metric and the
4D charged dilatonic black hole.  For these spherically symmetric
black holes in four dimensions, the dual backgrounds are presented
and explicitly shown to be new solutions of the leading order
string equations.  Some of these new backgrounds are found to
have no continuous isometries (except for time translations) and
also have naked singularities.}

\Date{10/92}


\lref\bn{I. Bars and D. Nemechansky, Nucl. Phys. B348 (1991) 89.}
\lref\witten{E. Witten, Phys. Rev. D44 (1991) 314.}
\lref\rwzw{E. Witten, Comm. Math. Phys. 92 (1984) 455.\semi
For a review see P. Goddard and D. Olive,
Int. J. Mod. Phys. 1 (1986) 303.}
\lref\wzwg{E. Witten, Nucl. Phys. B223 (1983) 422\semi
K. Bardacki, E. Rabinovici and B. Saering,
Nucl. Phys. B301 (1988) 151 \semi
D. Karabali and H. J. Schnitzer,
Nucl. Phys. B329 (1990) 649, and references therein.}
\lref\busc{T. Buscher, Phys. Lett. 194B (1987) 59;
 Phys. Lett. 201B (1988) 466.}
\lref\vhol{For a review see, J. W. van Holten, Z. Phys. C27:57 (1985).}
\lref\gmv{M Gasperini, J. Maharana and G. Veneziano, Phys. Lett. 272B (1991)
277.}
\lref\sen{A. Sen, Phys. Lett. 271B (1991) 295; Phys. Lett. 274B (1992) 34
\semi
S. Hassan and A. Sen, Nucl. Phys. B375 (1992) 103.}
\lref\md{M. Duff, talk at Trieste summer 1989,
Nucl. Phys. B335 (1990) 610.}
\lref\muller{M. Muller, Nucl. Phys. B337 (1990) 37.}
\lref\cfmp{C. Callan, D. Friedan, E. Martinec and M. Perry,  Nucl.
Phys. B262 (1985) 593.}
\lref\kir{E.B. Kiritsis, Mod. Phys. Lett. A6 (1991) 2871.}
\lref\vero{M. Ro\v cek and E. Verlinde, Nucl. Phys. B373  (1992) 630
.}
\lref\giva{P. Ginsparg and C. Vafa, Nucl. Phys. B289 (1987) 414 \semi
    T. Banks, M. Dine, H. Dykstra and
    W. Fischler, Phys. Lett. 212B (1988) 45\semi
    E. Alvarez and M. Osorio, Phys. Rev. D40 (1989) 1150.}
\lref\grvsw{A. Giveon, E. Rabinovici and G. Veneziano,
Nucl. Phys. B322 (1989) 167\semi
A. Shapere and F. Wilczek, Nucl. Phys. B320 (1989) 669.}
\lref\gz{M.K. Gaillard and B. Zumino, Nucl. Phys. B193 (1981) 221.}
\lref\cfg{S. Cecotti, S. Ferrara and Girardello,
Nucl. Phys. B308 (1988) 436.}
\lref\iltq{L.E. Ib\'a\~nez, D. L\"ust, F. Quevedo and S. Theisen
unpublished (1990) \semi G. Veneziano, Phys. Lett. B265 (1991) 287.}
\lref\bmq{C. Burgess, R. Myers and F. Quevedo unpublished (1991) \semi
A. Tseytlin, Mod. Phys. Lett. A6 (1991) 1721.}
\lref\bv{R. Brandenberger and C. Vafa, Nucl. Phys. B316 (1989) 391.}
\lref\mv{K. Meissner and G. Veneziano, Phys. Lett. 267B (1991) 33\semi
M. Gasperini and G. Veneziano, Phys. Lett. 277B (1992) 256.}
\lref\tv{A. Tseytlin and C. Vafa, Harvard preprint HUTP-91/A049
(hepth@xxx/\-9109048).}
\lref\dvv{R. Dijgkraaf, E. Verlinde, and H. Verlinde,
Nucl. Phys. B371 (1992) 269\semi
A. Giveon, Mod. Phys. Lett. A6 (1991) 2843.}
\lref\rAGG{L. Alvarez-Gaum\'e and P. Ginsparg,
Ann. Phys. 161 (1985) 423.}
\lref\rpglh{P. Ginsparg, ``Applied conformal field theory,''
Les Houches lectures (summer, 1988), published in
{\it Fields, Strings, and Critical Phenomena\/},
ed.\ by E. Br\'ezin and J. Zinn-Justin, North Holland (1989).}
\lref\rvafa{C. Vafa, ``Topological Mirrors and Quantum Rings'',
Harvard preprint HUTP-91/A059 (hepth@xxx/\-9111017).}
\lref\px{P. Candelas, X.C. de la Ossa, P.S. Green and
L. Parkes, Nucl. Phys. B359 (1991) 21; Phys. Lett. 258B (1991) 118.}
\lref\barst{I. Bars,
University of Southern California preprint USC-91/HEP-B4.}
\lref\hhs{J.H. Horne and G. Horowitz,
Nucl. Phys.  B368 (1992) 444
\semi J. Horne, G. Horowitz and A. Steif,
Phys. Rev. Lett. 68 (1992) 568.}
\lref\hhss{J. Horne, G. Horowitz and A. Steif,
University of California preprint
UCSBTH-91-53 (hepth@xxx/9110065).}
\lref\giro{A. Giveon and M. Ro\v cek, Nucl. Phys. B380 (1992) 123.}
\lref\ind {S.P. Khastgir and A. Kumar Bubaneswar, preprint
(hepth@xxx/9109026).}
\lref\cresc{M. Crescimanno, Berkeley preprint LBL-30947\semi
P. Ho\v rava Chicago preprint EFI-91-57 (hepth@xxx/9110067)\semi
I. Bars and K. Sfetsos, Univ. Southern California preprints
USC-91/HEP-B5 (hepth@xxx/\-9110054) and USC-91/HEP-B6
(hepth@xxx/\-9111040)
\semi D. Gershon, preprint TAUP-1937-91 (hepth@xxx/9202005).}
\lref\dlp{L. Dixon, J. Lykken and M. Peskin, Nucl. Phys. B325 (1989) 325.}
\lref\barsc{I. Bars, Nucl. Phys. B334 (1990) 125.}
\lref\hel{S. Helgason, ``Differential Geometry, Lie Groups, and Symmetric
spaces'', Academic Press (1978)\semi
R. Gilmore, ``Lie Groups, Lie Algebras and Some of Their
Applications'', Wiley (1974).}
\lref\host{G. Horowitz and A. Strominger,
Nucl. Phys. B360 (1991) 197.}
\lref\gique{P. Ginsparg and F. Quevedo, Nucl. Phys. B385 (1992) 527. }
\lref\einstein{C. Hoenselaers and W. Dietz (eds.) ``Solutions
of Einstein's Equations: Techniques and Results '', Springer Verlag,
Berlin (1984) .}
\lref\nicolai{H. Nicolai, ``Two--Dimensional Gravities and
Supergravities as Integrable Systems'', preprint DESY 91-038 (1991).}
\lref\narain{K. S. Narain, Phys. Lett. Bxxx, (1986)}
\lref\flst{S. Ferrara, D. L\"ust, A. Shapere and S. Theisen, Phys. Lett.
B225 (1989) 363\semi E. J. Chun, J. Mas, J. Lauer and H.P. Nilles,
Phys. Lett. B233 (1989) 141\semi M. Cveti\v c, A. Font, L.E. Ib\'a\~nez,
D. L\"ust and F. Quevedo, Nucl. Phys. B361 (1991) 194.}
\lref\bigs{S. Weinberg, Field Theory Notes (1984).}
\lref\bigsd{See for instance, S. Weinberg, ``Gravitation and Cosmology'',
Wiley (1972).}
\lref\bars{I. Bars and K. Sfetsos, Phys. Lett. 277B (1992) 269; Mod. Phys.
Lett. A7 (1992) 109; USC preprints USC-92-HEP-B1, B2.}
\lref\gibbons{G.W. Gibbons and K. Maeda, Nucl. Phys. B298 (1988) 741\semi
  D. Garfinkle, G. Horowitz and A. Strominger, Phys. Rev. D43 (1991) 3140.}
\lref\duality{K. Kikkawa and M. Yamasaki, Phys. Lett. 149B (1984) 357\semi
N. Sakai and I. Senda, Progr. Theor. Phys. Suppl. 75 (1986) 692.}
\lref\tseyt{A. Tseytlin, ``Cosmological Solutions with Dilaton and
Maximally Symmetric Space in String Theory'', Cambridge preprint
DAMTP--15--1992 (1992). }
\lref\nappi{C. R. Nappi, Phys. Rev. D21 (1980) 418\semi
     B. E. Fridling and A. Jevicki, Phys. Lett. 134B (1983) 70\semi
     E. S. Fradkin and A. A. Tseytlin, Ann. Phys. 162 (1985) 48.}
\lref\tseytd{A. A. Tseytlin, Mod. Phys. Lett. A6 (1991) 1721\semi
            A. S. Schwarz and A. A. Tseytlin, preprint
	    Imperial/TP/92-93/01.}
\lref\savit{R.Savit, Rev. Mod. Phys. 52 (1980) 453.}
\vskip.6in

\newsec{Introduction}

One of the most interesting properties discovered in
nontrivial string backgrounds is the existence of target space
duality \duality. Besides providing a
better understanding of the  moduli space of a given solution,
it may lead to interesting cosmological consequences as emphasized
in \bv. It can also
 provide crucial information about the low energy
couplings in  string compactifications, by requiring the interactions
to be expressed in terms of the modular functions of a `duality'
group \flst .

Duality symmetries were originally
discovered for toroidal compactifications
of  closed string theories \duality. Subsequently,
it has been realized that they are a
property of all string vacua with abelian isometries \busc\
and the invariance of the partition function under
a duality transformation was shown for the corresponding
$\s$--model.
In this way, duality on curved string backgrounds such as
2D black holes  or more general gauged WZW
models has been understood \refs{\dvv,\hhs, \kir, \gique}.
This however cannot be the end of the story since
similar symmetries exist for compactifications on
Calabi--Yau spaces \px , even though these are string
backgrounds without isometries\footnote{$^*$}{Duality symmetries
for Calabi--Yau compactifications are analogous
to duality symmetries for toroidal compactifications.
They are the symmetries
in the modular group of the K\"ahler class parameter space of
the Calabi--Yau metric and relate ``large'' with
``small'' manifolds.}.  At the moment
there is no understanding of these symmetries
in terms of the $\s$--model action.

In this paper we address the question of whether there
exist duality transformations
for string backgrounds with non--abelian isometries.
The physical motivation is clear.  The discovery of
these duality symmetries
represents another step in the classification of physically
inequivalent string vacua.  It is for
example very interesting
to study solutions of Einstein's equations in vacuum since
they are also solutions of the leading order string
background equations with constant dilaton and antisymmetric
tensor field.
It also happens that  {\it every} known solution
of Einstein's equations in 4D has isometries \einstein.
Some of the
most interesting 4D geometries,
such as the Schwarzschild solution
and the Friedman--Robertson--Walker (FRW) homogeneous
cosmologies,
have non--abelian isometries: they have an
$SO(3)$ spherical symmetry.

Given a $\s$--model action with a global symmetry,
the procedure we follow starts by gauging the symmetry in the
$\s$--model action and adding to it an extra term
with a Lagrange multiplier which constrains the gauge field
strength to vanish.
After integrating over the Lagrange
multiplier and fixing the gauge, we recover the original action.
On the other hand, by
integrating by parts the Lagrange multiplier term and then
integrating out the gauge fields, we obtain the dual action in which
the
Lagrange multiplier is a new dynamical field. This procedure
is equivalent
to the standard first order formalism used for abelian
isometries \refs{\busc, \gique} as emphasized recently in
\refs{\vero,\giro}, but it generalizes more readily
to the non--abelian case\footnote{$^*$}{For a related discussion
of duality in terms of the exchange of Bianchi identities and field
equations of 2D $\s$--models on group manifolds, see \nappi.  For a
review of earlier work on duality in field theories with non--abelian
symmetries see \savit.}.
As we will see, the question of whether there exist
duality symmetries, can be posed for any worldsheet
$\s$--model with target space isometries, regardless of the
conformal invariance of the theory.  Conformal invariance
is however mantained by also transforming the dilaton
field appropriately \refs{\giva,\busc}.

Probably the most intriguing property of this new duality that
we have found is that
it can map a geometry with non--abelian isometries to another which
has none. This is remarkable because starting from
the geometry with no isometries, the current
procedure for performing duality transformations would not give any
information about the existence of the `dual'
geometry.
Nevertheless the two geometries are indeed related and even though
they are very different as geometries, they give the same
partition function. This indicates
that there should be a more general argument, deeper
and independent of
the existence of continuous isometries,
by means of which duality transformations can be explained. As
mentioned above, this is also implied by the existence of
duality--like symmetries in Calabi--Yau spaces.

To set up conventions and notation, as well as to make the
paper self contained, in Sect. 2 we review briefly the duality
transformation for the case of abelian isometries
of the target space.  In Sect. 3
we present the generalization to non-abelian
isometries.  We discuss in detail the change in the
dilaton field necessary to render the theory
conformally invariant to first order.
We give the duality transformation for $SO(N)$
and present it explicitly for $SO(3)$, which is the
case relevant to 4D spherically symmetric solutions.
In Sect. 4 we find the dual geometries of Schwarzschild
and of
charged dilatonic black holes, for which the isometry
group is $SO(3)\times\{ {\rm time\ translations}\}$.  We make
concluding remarks
and discuss our results in Sect. 5.

\newsec{Duality with Respect to Abelian Isometries}

Target space duality is a general property of string
vacua which have at least one
isometry. We will briefly review here the case where the
isometries are abelian in order to be able to understand
the generalization to non--abelian isometries.
In ref.~\busc, the `$r\to 1/r$' duality in toroidal compactifications
of string theories was generalized to any string background
for which the metric in the worldsheet action had at least one
isometry.  Examples where this duality transformation
has been applied are
the 2D black-hole and some simple cosmological string
backgrounds. The worldsheet action for the bosonic string in a
background with $N$ commuting isometries, can be written as
\eqn\sigmad{\eqalign{S = {1\over{4\pi\alpha'}}\int &\d^2z\,
\Bigl(Q_{\mu\nu}(X_{\alpha})\,\del X^{\mu} \delbar X^{\nu}
 + Q_{\mu n}(X_{\alpha})\del X^{\mu} \delbar X^n \cr
 &\quad + Q_{n \mu}(X_\alpha)\del X^n \delbar X^{\mu}
 + Q_{mn}(X_\alpha)\del X^m \delbar X^n \cr
 &\quad + \alpha' R^{(2)}\Phi(X_\alpha)\Bigr)\ ,\cr}}
where $Q_{MN}\equiv G_{MN}+B_{MN}$ and lower case
latin indices $m,n$ label the isometry directions. Since the
action \sigmad\ depends on the $X^m$
only through their derivatives, we can write it in
first order form by introducing
variables $A^m$ and adding an extra term to the action
\hbox{$\L_m(\del \overline A^m - \delbar A^m)$} which imposes
the constraint $A^m=\del X^m$.
Integrating over the Lagrange multipliers $\L_m$ returns
us to the original action
\sigmad. On the other hand
performing partial integration and solving for
$A^m$ and $\overline A^m$, we find the dual action $S'$
which has an identical form to $S$ but with
the dual background given by \refs{\busc,\gique}
\eqn\qp{\eqalign{Q'_{mn}& = (Q\inv)_{mn}\cr
  Q'_{\mu \nu}& = Q_{\mu \nu} - Q_{\mu m}\,(Q\inv)^{mn}\,Q_{n\nu}\cr
  Q'_{n \mu}& = (Q\inv)_n^{\ m}\, Q_{m \mu}\cr
  Q'_{\mu n}& = - Q_{\mu m}\,(Q\inv)^m_{\ n}\ .\cr}}
To preserve conformal invariance, it can be seen  \refs{\busc,
\giva} that the dilaton field has to transform as
$\Phi' = \Phi - \log\det G_{mn}$.
Notice that equations \qp\ reduce to the usual
duality transformations for the toroidal compactifications of
\grvsw\ in the case $Q_{m\mu}=Q_{\mu m}=0$ and can map
 a space with no
torsion ($Q_{m\mu} =Q_{\mu m}$)  to a space
with torsion ($Q'_{m\mu}=-Q'_{\mu m}$).
 For
the case of a single isometry, we recover the
explicit expressions of \busc.

It is not necessary to go to the
first order formalism for every isometry direction
since we do not have to perform a duality
transformation for all of them.  That is, we can
integrate over a subset of the
Lagrange multipliers $\L_m$ and, for the remaining
isometry directions, integrate out
the corresponding gauge
fields $A^m$.
Equation \qp\  should then be read with the
indices $m,n$ running only over
the variables with isometries that have been dualized. The
total duality group includes these transformations
as well as shifts in the antisymmetric tensor field and
has been argued \refs{\cfg, \md, \giro} to be equivalent
to $SO(N,N,\IZ)$.

An equivalent interpretation of the duality process
just described is given in \vero. In the original action
the symmetry is gauged by replacing
$\del X^m$ with $D X^m = \del X^m + A^m$ and the
term $\int d^2z\ \L_m(\del \overline A^m - \delbar A^m)$
is added to the action.  This extra term
imposes the vanishing of the field strength $F$
of the gauge fields after integration over the
Lagrange multipliers $\L_m$.
This implies that the gauge field must be pure gauge,
$A^m = \del\widetilde X^m$.
The gauge fixing can be done either by
choosing the gauge fields to vanish or by taking
$X^m=0$ (a unitary gauge). In both cases this
reproduces the original action.
The dual theory is obtained by instead integrating out the
gauge fields and then fixing the gauge.
This is the procedure that generalizes
the duality transformation to the case
of non--abelian isometries.

\newsec{Duality with Respect to Non--Abelian Isometries}

Consider  the $\sigma$--model action \sigmad\ and
assume that the target space metric has a group $\cal{G}$ of
non--abelian isometries.  In this case, $Q_{MN}$ {\it does}
depend on $X^m$ and transforms accordingly
under $X^m\to g^m_{\ n} X^n\ , g\in \cal{G}$.
We now follow the procedure
of Ref\vero.  We gauge the symmetry corresponding to a
subgroup $H\subseteq\cal{G}$
\eqn\dtoD{
    \del X^m \to DX^m = \del X^m +
              A^{\alpha}(T_{\alpha})^m_{\ n} X^n\ ,}
and add to the action the term
\eqn\constraint{\int d^2 z\ tr(\L F) =
                        \int d^2 z\ \L_{\alpha}F^{\alpha}\ ,}
where in this case the gauge field strength is,
in matrix notation,
\eqn\nonaF{F =
       \del \overline A - \delbar A + [A,\overline A]\ .}
The  $N \times N$ matrices $T_{\alpha}$ form an
adjoint representation of the group $H$,
$[T_{\alpha},T_{\beta}] =
         c_{\alpha\beta}^{\quad\gamma} T_{\gamma}\ $,
normalized such that
$tr(T_{\alpha} T_{\beta}) = \delta_{\alpha\beta} \ $
(a constant in the normalization can be absorbed in a
redefinition of the Lagrange multiplier ).
In the path integral we have then
\eqn\pathint{\eqalign{
   {\cal P} &= \int {\cal D}X\  e^{- iS[X]}\cr
    &= \int {\cal D}X
         \int D\L \int {{DA D\overline A}\over {V_{\cal G}}}\
       \exp\left\{ -i\left( S_{gauged}[X,A,\overline A] +
                 \int d^2 z tr(\L F) \right) \right\} \ ,\cr}}
where $V_{\cal G}$ is the ``volume'' of the group of isometries
and ${\cal D}X$ is the measure that gives the correct volume
element
\eqn\meas{{\cal D}X = DX \sqrt{G} e^{-\Phi}\ .}

Similar to the abelian case, the original action is obtained by
integrating out the Lagrange multiplier $\L$.  Locally, this forces
the gauge field to be pure gauge
\eqn\puregauge{A = h^{-1}\del h\ ,
    \quad \overline A = h^{-1}\delbar h\ ,\qquad h\in H\ .}
By fixing the gauge with the choice
$A = 0$, $\overline A = 0$ we reproduce
the original theory.  A different gauge choice will only give
the same theory in a different coordinate system.

The dual theory is obtained by integrating out
the gauge fields in the path integral \pathint.
Integrating over the gauge fields $\overline A$ we obtain
\eqn\pathintt{
{\cal P} = \int {{\cal D}X\over V_{\cal G}}\ D\L
          \int DA\ \delta\left( fA + h \right)\
       \exp\left\{- i\left( S[X] +
    {1\over{4\pi\alpha'}}\int d^2 z\
       \overline h_{\alpha} A^{\alpha}\right)\right\}\ ,}
where $S[X]$ is the original action and $h$, $\overline h$
and the matrix $f$ are given by
\eqn\fhbarh{\eqalign{
  h_{\alpha} &= - \del \L_{\alpha} +
        \left(Q_{\mu n}\del X^{\mu} +
              Q_{kn}\del X^k\right) (T_{\alpha})^n_{\ m} X^m\ ,\cr
  \overline h_{\alpha} &= \- \delbar \L_{\alpha} +
        \left(Q_{n \mu }\delbar X^{\mu} +
              Q_{nk}\delbar X^k\right) (T_{\alpha})^n_{\ m} X^m\ ,\cr
  f_{\alpha \beta} &= - c_{\alpha \beta}^{\quad\gamma} \L_{\gamma} +
     X^k (T_{\beta})^j_{\ k} Q_{jn} (T_{\alpha})^n_{\ m} X^m \ .\cr}}
Integrating out the gauge field $A$ we thus obtain
\eqn\predualPI{{\cal P} =
               \int {{\cal D}X\over V_{\cal G}}\ D\L\
                              e^{-iS'[X,\L]} \det (f^{-1})\ ,}
with $S'[X,\L]$ given by
\eqn\predualS{S'[X,\L] = S[X]
               - {1\over{4\pi\alpha'}}\int d^2 z\
            \overline h_{\alpha} (f^{-1})^{\alpha\beta} h_{\beta}\ .}

We should now fix the gauge to eliminate
the extra degrees of freedom that were introduced by
gauging the original action.  To do this we use
the gauge symmetry to introduce a convenient gauge:
$\widehat X = h_0 X$,
$\hat \L = h_0^{-1} \L h_0$, for a given $h_0\in H$.  Obviously,
different
gauge choices {\it will not} give different dual theories,
they will give the same theory differing by a
coordinate transformation.  Notice that the maximum
number of Lagrange multipliers that can be gauged
away is ${\rm dim}{\cal G} - {\rm rank}{\cal G}$.  Indeed,
the number of Lagrange multipliers invariant under
the group of isometries is ${\rm rank}{\cal G}$.
Using the Fadeev--Popov method to fix the
gauge in the path integral we obtain
\eqn\ppredualPI{{\cal P} =
               \int {\cal D}X\ D\L\
	 \delta[{\cal F}]\ \det{{\delta{\cal F}}\over{\delta\omega}}\
                              e^{-iS'[X,\L]} \det (f^{-1})\ ,}
where ${\cal F}$ is the gauge fixing function and $\omega$ are the
parameters of the group of isometries.  Therefore
\eqn\dualPI{{\cal P} =
        \int {\cal D}Y\ e^{-iS'[Y]} \det (f(Y)^{-1}) \ .}
Here we have denoted the new coordinates
$\widehat X$ and $\hat\Lambda$ on the dual manifold
collectively by $Y$.  The Fadeev--Popov determinant
in the path integral contributes to the measure such that
the correct volume element for the dual manifold is {\it precisely}
obtained
\eqn\dualmeas{{\cal D}Y = DY \sqrt{G'} e^{-\Phi'}\ .}
The factor $\det (f^{-1})$ in the partition function
can be computed using standard heat kernel
regularization techniques
(see  for example references \busc, \kir\ and \tseytd). It generates
a new local term in the action of the form
\eqn\dilterm{{1\over{4\pi\alpha'}}\int d^2 z\
         \alpha' R^{(2)}\ (\Delta\Phi )\ ,}
which corresponds to the change in the dilaton
due to the duality transformation
\eqn\newdil{\Phi' = \Phi - \log\det f\ .}
This change in the dilaton transformation is the
shift necessary to retain the conformal invariance
of the dual theory.  There is actually another prescription
from which this change in the dilaton can be obtained.  In fact,
the requirement that the correct volume element \dualmeas\
is obtained in the dual theory means that
\eqn\newdila{e^{-\Phi'}= \left[ e^{-\Phi}\ \sqrt{G\over {G'}}\
             \det {{\delta {\cal F}}\over {\delta\omega}}
	        \right]_{{\cal F}=0}\ ,}
as can be checked from equations \ppredualPI\ and \dualmeas.
This prescription coincides with the prescription
for duality with respect to abelian isometries in
\busc\ because in this case
the Fadeev--Popov determinant is trivial.  A consistency
check of the change in the dilaton is obtained by comparing
equations \newdila\ and \newdil.

In general, we cannot write explicitly the gauge fixed
dual action. Therefore, we are not able to
present the new metric and antisymmetric tensor fields in a closed
form, as was done for the abelian case in equations \qp.
In the following we shall study a number of examples
and in these cases the gauge fixing procedure
will be carried through in detail. This will allow us to
have explicit expressions for the dual background fields.
As an example to which we will extensively refer, let us
consider a theory  for which the target space metric
has a maximally symmetric subspace
with ${\cal G} = SO(N)$ and no
antisymmetric tensor. The coordinates $X^M, M = 1,...,D$,
can be decomposed into $N-1$ angular coordinates
($\theta^i$) describing
$(N-1)$--dimensional spheres, and $D-N+1$ extra
coordinates ($v^\mu$) specifying the different spheres in
the $D$ dimensional spacetime. The metric can then be
decomposed as \bigsd\ in the form
\eqn\split{ ds^2= g_{\mu\nu}(v)dv^\mu dv^\nu
+\Omega (v) g_{ij} d\theta^i  d\theta^j\ .}
It is more convenient to
treat the coordinates $\theta^i$ in terms
of cartesian coordinates $X^m$ in
a $N$ dimensional space on which $SO(N)$ can
act linearly, so
we write the $\s$ model action in the form
\eqn\sonS{\eqalign{S[v,X] &= S[v]+\int d^2 z\ \Omega(v)
      \left\{ g_{mn}\del X^m \delbar X^n +
  {1\over{2a\sqrt{\Omega}}}\lambda (g_{mn} X^m X^n - a^2)\right\}\cr
     &+ {1\over{4\pi\alpha'}}\int d^2 z\
         \alpha' R^{(2)}\ \Phi\ ,\cr}}
where
$S[v] = \int d^2 z g_{\mu\nu}(v)\del v^{\mu} \delbar v^{\nu}$,
the metric $g_{mn}$ is
diagonal and constant and
the Lagrange multiplier term fixes
the $N$ dimensional space to be
a sphere of radius $a$.  The factor ${1\over{2a\sqrt{\Omega}}}$
has been introduced to obtain the correct volume element of
the sphere after integrating over $\lambda$.
Gauging this action (with a vanishing field strength)
and fixing the gauge $A=\bar A=0$ we obtain
the original action.
To find the dual action we can take antisymmetric
matrices $T_{\alpha}$
and $h$, $\overline h$ and $f$ are given by
\eqn\prefhhson{\eqalign{
  h_{\alpha} &= - \del \L_{\alpha} +
            \Omega(v)\del X^n (T_{\alpha})_{nm} X^m\ ,\cr
  \overline h_{\alpha} &= \- \delbar \L_{\alpha} +
            \Omega(v)\delbar X^n (T_{\alpha})_{nm} X^m\ ,\cr
  f_{\alpha\beta} &= - c_{\alpha\beta}^{\quad\gamma} \L_{\gamma} -
     {1\over 2} \Omega(v)
         X^n \left\{ T_{\beta},T_{\alpha}\right\}_{nm} X^m \ .\cr}}
A convenient choice of gauge is to set
\eqn\rgauge{X^m = 0,\quad m = 1,..., N-1,\quad X^N = a}
and then use the remaining $SO(N-1)$ gauge freedom (the gauge
transformations that preserve \rgauge)
to gauge away ${1\over2} (N - 1) (N - 2)$ of the Lagrange
multipliers $\L_{\alpha}$.  We then obtain
\eqn\fhhson{\eqalign{
  h_{\alpha} &= - \del \L_{\alpha}\ ,\cr
  \overline h_{\alpha} &= \- \delbar \L_{\alpha}\ ,\cr
  f_{\alpha \beta} &= - c_{\alpha \beta}^{\quad\gamma} \L_{\gamma}
             - a^2 \Omega(v) (T_{\alpha})_{Nm}(T_{\beta})^m_{\ N}\ ,\cr}}
and
\eqn\dualsons{S^{dual}[v,\L] = S[v]+{1\over{4\pi\alpha'}}
                 \int d^2 z\ \left(
               \del\L_{\alpha} (f^{-1})^{\alpha\beta} \delbar\L_{\beta}
                   \right) +
          {1\over{4\pi}}\int d^2 z\ R^{(2)}\Phi' \ .}
{}From this expression,  we can in principle read off
the new background fields
as in \qp. The only problem is that we  still  have to complete
the gauge fixing for the $\L_{\alpha}$. In order to
 be more explicit,
we will now consider  in greater detail examples with
${\cal G} =SO(2)$ and
${\cal G} =SO(3)$.  The case for $SO(2)$, even
though abelian, will be
considered to fix ideas and to show how the formalism contains the
abelian case.

When ${\cal G}=SO(2)$ there is only one matrix $T$
\eqn\Tsotwo{T = \pmatrix{\- 0 &\- 1\cr -1 &\- 0\cr}\ .}
In the gauge $X^1 = a$, $X^2 = 0$ and $A=\del\theta$ (that is,
in spherical coordinates), the original action
\sonS\  reduces to
\eqn\sotwoS{S[v,\theta] = S[v] +
     \int \d^2 z\ a^2 \Omega(v)\del\theta\delbar\theta\ .}
To obtain the dual theory we start by calculating the
quantities $h$, $\overline h$ and $f$.  Before
fixing the gauge we have
\eqn\prefhhsotwo{\eqalign{ f &= a^2\Omega(v)\ ,\cr
        h &= -\del\Lambda - a^2\Omega(v) \del\theta\ ,\cr
 \overline h &= \-\delbar\Lambda - a^2\Omega(v)\delbar\theta\ ,\cr}}
where we have used spherical coordinates
%
$X^1 = a \cos\theta\ , X^2 = a \sin\theta\ $.
The action \predualS\ before fixing the gauge is therefore
\eqn\presotwos{S'[v,\L] = S[v]+{1\over{4\pi\alpha '}}\int d^2 z\
   \left( {1\over {a^2\Omega(v)}} \del\L\delbar\L +
     \del\theta\delbar\L - \del\L\delbar\theta\right)\ .}
After fixing the gauge by choosing $\theta=0$, we obtain
for the dual action
\eqn\dualsotwos{S^{dual}[v,\L] =
    S[v]+{1\over {4\pi\alpha '}}\int d^2 z\
       \left({1\over {a^2\Omega(v)}} \del\L\delbar\L\right) +
  {1\over {4\pi}} \int d^2 z\ R^{(2)} \Phi ' \ ,}
where $\Phi' = \Phi - \log a^2\Omega(v)$.  As expected, we have
recovered the `$r\to {1\over r}$' duality symmetry on the circle.
It is important
to note here that when fixing the gauge
we could not have gauged away
the Lagrange multiplier $\L$  instead of
eliminating $\theta$ because it is {\it invariant} under
the SO(2) gauge transformations.  In the dual theory, what
started life as a Lagrange multiplier became a coordinate.

Note also that the dual metric \dualsotwos\ has an isometry since it
depends on $\L$ only through its derivatives.  This feature is general:
if a metric has a commutative isometry associated to a
coordinate $\varphi$, then the dual
theory with respect to this isometry also has an isometry
associated to the coordinate $\L$, dual to $\varphi$, which
appeared as a Lagrange multiplier in passing to
the dual theory.  This is easily seen by noting that
when the isometry is abelian  the term \constraint\ is invariant
under constant shifts of $\L$ as a consequence of partial
integration.  This
symmetry is mantained even after integrating out the
gauge fields.  In fact, when the isometry is abelian,
the matrix $f_{\alpha\beta}$ in \prefhhson\ becomes
completely independent of $\L$ since the structure
constants vanish and hence,
the dual action depends on $\L$ only through its derivatives.
It is important to remark though that abelian isometries
disappear under duality with respect to a group of
non--abelian isometries that contains the corresponding abelian
group.

We would like to remark that this formalism works also in
coordinate systems in which the abelian isometries
do not manifest themselves as constant shifts of the
coordinates.  Consider for example the 2D black hole
\witten\ with metric in Kruskal--like coordinates
\eqn\twodbh{ds^2 = - { du dv\over {1 - uv}}\ .}
The corresponding nonlinear $\sigma$--model does
not have the form \sigmad\ because the metric
does depend explicitly on $u$ and $v$.  To perform
a duality transformation following \busc\ is therefore
necessary to change coordinates so that the
$\sigma$--model has the form \sigmad.  However, since the
isometry of \twodbh\ manifests itself as a SO(1,1)
symmetry action on the coordinates ($u\to e^{\alpha}u$ and
$v\to e^{-\alpha}v$), we can follow our formalism
to find the dual metric without performing first
a coordinate transformation.  We reproduce the fact that
this background is self dual \dvv.

For ${\cal G} = SO(3)$ there are three matrices $T$.  Due to the
fact that in three dimensions an antisymmetric matrix is
dual to a three--vector,
we can choose $(T^p)^m_{\ n} = \epsilon^{pm}_{\quad n}\ $.
The original action in spherical coordinates (the gauge choice here
is $X^1 = X^2 = 0$, $A^1 = \del\theta$,
$A^2 = - \sin\theta\del\varphi$ and
$A^3 = \cos\theta\del\varphi$) is
\eqn\sots{S[v,\theta,\varphi] = S[v] + \int d^2 z\ a^2\Omega(v)
  \left(\del\theta \delbar\theta +
          \sin^2\theta \del\varphi\delbar\varphi
                        \right)\ .}
We now find the dual theory.  Before fixing the gauge
\eqn\fhhsothree{\eqalign{
      h_p &= -\del\L_p + \Omega(v)\del X^m \epsilon_{pmn} X^n\ ,\cr
   \overline h_p &= \-\delbar\L_p +
               \Omega(v)\delbar X^m \epsilon_{pmn} X^n\ ,\cr
  f_{pq} &= \epsilon_{pq}^{\quad m} \L_m +
      \Omega(v)\left( \delta_{pq} a^2 -
      \delta_{pm} X^m \delta_{qj} X^j\right)\ .}}
Fixing the gauge as in \rgauge, choosing $\L_2 = 0$ (we are not able
to gauge away $\L_3$ once we have chosen \rgauge) and defining
$ x^2 = \L_1^{\ 2} + \L_3^{\ 2}$ and $y = \L_3$, we obtain
the dual theory action
\eqn\dualsothrees{\eqalign{
    S^{dual}[v,x,y] = & S[v]+{1\over {4\pi\alpha '}} \int d^2 z\
          {1\over {a^2\Omega(v) (x^2 - y^2)}}\
    \left( a^4 \Omega(v)^2\del y \delbar y
                    + x^2 \del x \delbar x \right) \cr
  &\qquad + {1\over {4\pi}} \int d^2 z\ R^{(2)}\Phi'\ ,\cr}}
where
\eqn\dilsothree{\Phi ' = \Phi - \log [a^2 \Omega(v)\ (x^2 - y^2)]\ .}
To understand this metric better, we pass to
new coordinates
$x = \lambda$, $y = \lambda\cos\theta$.
This coordinate system corresponds to the gauge choice
$X^1 = 0$, $X^2 = a \sin\theta$,
$X^3 = a \cos\theta$, $\L_1 = \L_2 = 0$
and $\lambda = \L_3$.
The dual action is now
\eqn\ddualsothree{\eqalign{S^{dual}[v,\theta,\lambda] &= S[v]\cr
  & +{1\over {4\pi\alpha '}} \int d^2 z
    \Biggl\{a^2\Omega(v) \left( \del\theta -
               \cot\theta{{\del\lambda}\over\lambda}\right)
                        \left( \delbar\theta -
               \cot\theta{{\delbar\lambda}\over\lambda}\right)\cr
	       & \qquad\qquad\qquad\qquad +
               {1\over {a^2\Omega(v) \sin^2\theta}}
		    \del\lambda\delbar\lambda\Biggr\}\cr
  & + {1\over {4\pi}} \int d^2 z\ R^{(2)}\ {\Phi'}\ ,\cr}}
with
\eqn\ddilsothree{\Phi' = \Phi -
          \log[a^2 \Omega(v)\ \lambda^2\sin^2\theta]\ .}
It can be seen from \ddualsothree\ that a duality
transformation with respect to $SO(3)$ of the round metric on a
two sphere, gives the metric that we would have obtained if
we had made a duality transformation only with respect to
the $U(1)$ subgroup of $SO(3)$ which represents the invariance
of the original metric under
$\varphi\to\varphi+ {\rm constant}$, except for
the fact that under duality with respect to $SO(3)$
\eqn\twist{d\theta\to d\theta
               - \cot\theta {{d\lambda}\over\lambda}\ .}
In fact, the dual theory to \sonS\ for $N=3$ with respect to the
abelian isometry  {\hbox{$\varphi\to\varphi+ {\rm constant}$}} is
\eqn\dualUone{\eqalign{S^{dual}[v,\theta,\lambda] = S[v] &+
   {1\over {4\pi\alpha '}} \int d^2 z\
      \left( a^2 \Omega(v)\del\theta\delbar\theta +
        {1\over {a^2 \Omega(v)\sin^2\theta}}\ \del\lambda\delbar\lambda
             \right) \cr
 &+ {1\over {4\pi}} \int d^2 z\ R^{(2)} \Phi' \ ,\cr}}
where the new dilaton in this case is
\eqn\ddiluone{\Phi'= \Phi - \log[a^2 \Omega(v)\ \sin^2\theta]\ .}

We can now ask the question of whether the isometries of the
original metrics are preserved.  The answer is that generically
this is {\it not} the case, that is, the original
group of isomeries ${\cal G}$ is broken by the
duality transformation.  In our example for
$SO(3)$ the original metric had three isometries.
However \dualUone\ has only one and \ddualsothree\ has none
as can be shown by solving directly
the Killing equations.  For the coordinate
transformations $X^m\to X^m + \epsilon\zeta^m$ and $v^{\mu}$
invariant, the Killing equations in these metrics are
\eqn\kill{\eqalign{0 &= \pd{\zeta^m}{v^{\nu}}g_{mn}\ ,\cr
  0 &= \pd{\zeta^m}{X^p}g_{mn} + \pd{\zeta^m}{X^n}g_{mp}
            + \zeta^m \pd{g_{pn}}{X^m}\ .\cr}}
The first equation means
that the Killing vectors $\zeta^m$ do not depend on the
coordinates $v^{\mu}$.
For the metric \dualUone\ the second equation gives
\eqn\killUone{\eqalign{
  0 &= \pd{\zeta^{\theta}}{\theta}\ ,\cr
  0 &= {1\over{a^2 \Omega\ \sin^2\theta}}\pd{\zeta^{\lambda}}{\theta}
                + a^2 \Omega\ \pd{\zeta^{\theta}}{\lambda}\ ,\cr
  0 &= \pd{\zeta^{\lambda}}{\lambda}
      - \cot\theta \zeta^{\theta}\ ,\cr}}
which have the unique solution
\eqn\killvUonee{\eqalign{\zeta^{\theta} &= 0\ ,\cr
                         \zeta^{\lambda} &= {\rm constant} \ .\cr}}
That is, the only Killing vector of the metric in
\dualUone\ is the Killing vector corresponding to
constant shifts of $\lambda$.  The Killing
equations for the metric in \ddualsothree\
can also be solved explicitly since they are not much more
complicated than equations \killUone.  Actually, it is easier
to do the calculation in the metric in \dualsothrees\
since it its diagonal.  The Killing equations are in this
case
\eqn\killsothree{\eqalign{
  0 &= x \pd{\zeta^x}{x}
         - {y^2 \over {(x^2 - y^2)}} \zeta^x
         + {{x y} \over {(x^2 - y^2)}} \zeta^y\ ,\cr
  0 &= a^2 \Omega\ \pd{\zeta^y}{x}
         + {x^2 \over {a^2 \Omega}} \pd{\zeta^x}{y}\ ,\cr
  0 &=  \pd{\zeta^y}{y}
         - {x \over {(x^2 - y^2)}} \zeta^x
         + {y \over {(x^2 - y^2)}} \zeta^y\ ,\cr}}
and it is easy to see that they have no solution.
Therefore the metric in \dualsothrees\ (or \ddualsothree)
has no continuous isometries (except of course for possible isometries
of the metric $g_{\mu\nu}$ which have not been affected
by the duality transformation with respect to the
coordinates $X^m$). Notice however that  \dualsothrees\ has the $\IZ_2$
discrete isometries $x\to-x, y\to -y$ and $x\to x, y\to -y$
. These
symmetries belong to the original $O(3)$ transformations which are not
connected to the identity and are left untouched by the proce
ss
of dualization.

\newsec{Dual Geometries of 4-D Black Holes}

We will now present, as a matter of illustration, some 4D black hole
backgrounds
and their duals. In order for the dual
geometries to give  string vacua,
these geometries have to satisfy the string background
equations.
To lowest order in $\alpha '$ these equations are \cfmp\
\eqn\einst{R_{MN} + D_M D_N \Phi -
            {1\over 4}H_M^{LP} H_{NLP}=0}
\eqn\hmn{D_L H^L_{MN}-(D_L\Phi)H^L_{MN}=0}
\eqn\dil{R-2\Lambda-(D\Phi )^2 +
            2D_M D^M\Phi-{1\over 12}H_{MNP}H^{MNP}=0\ ,}
where $\L\equiv (c-26)/3$ is the cosmological constant in the
effective string action, $c$ is the central charge and, as usual
$H_{MNP}\equiv \del_{[M} B_{NP]}$.  For the heterotic
string, these equations
have to be modified in order to include the background gauge fields.

Since any solution of Einstein's equations in vacuum is
also a solution of \einst--\dil\  for constant dilaton
$\Phi$ and antisymmetric tensor $B_{MN}$, we have  at
hand large classes of solutions
of \einst--\dil\ with isometries \einstein. In
particular the Schwarzschild
metric
\eqn\schwars{ds^2=-(1-{2M/r}) dt^2 + {dr^2\over 1-{2M/r}}
                    + r^2(d\theta^2 + \sin^2\theta d\phi^2)\ ,}
times any CFT with $c= 22$ is a solution of \einst--\dil\
($c=6$ CFT is needed for the heterotic string,
which can be obtained either by
toroidal compactifications or Calabi--Yau spaces). The isometry group
is given by time translations $t\to t+t_0$ together
with the $SO(3)$ space rotations.

Direct application of the standard duality
transformation to \schwars\ for time
translations, gives the dual metric
\eqn\schwarsdu{ds^2=-{dt^2\over 1-{2M/r}} + {dr^2\over 1-{2M/r}}
                 + r^2(d\theta^2 + \sin^2\theta d\phi^2)\ ,}
with the dilaton field now given by $\Phi'=\Phi - \log(1-{2M/r})$.
This metric defines a geometry with naked singularities at
$r=0$ and $r=2M$, as it can be verified by
computing the curvature scalar
$R={4M^2\over (2M-r)r^3}$.  It is easy to check
that equations	\einst--\dil\  are satisfied by
the dual metric and dilaton $\Phi'$. We have then
found a spherically symmetric solution of the
string background equations, which is not a black hole,
but has naked singularities  and
is dual to the Schwarzschild solution.

It is interesting to find the dual of the Schwarzschild
black hole directly in Kruskal coordinates.  This is done as
for the 2D case described in the previous section.
The dual geometry in these coordinates has the same
singularity structure as in the dual of the
Schwarzschild metric.
This is different to the 2D case which is self
dual.  Therefore, even though the Kruskal diagrams
for the 2D and 4D black holes are identical, their
dual geometries are completely different.

Consider now the dual geometry of \schwars\ with respect to the
$SO(3)$ symmetry.  We find
\eqn\schwarsdd{ds^2=-(1-{2M/r}) dt^2 + {dr^2\over 1-{2M/r}}
   +{1\over {r^2 (x^2 - y^2)}}\ [r^4 dy^2
                    + x^2 dx^2]\ ,}
with the new dilaton $\Phi' = \Phi - \log [r^2 (x^2 - y^2)]$. The
regions $x=y$
and $r=0$ are real singularities whereas
$r=2M$ is only a metric singularity
corresponding to a horizon as in the original case. In fact, the
curvature scalar is
\eqn\curvs{R= -{(2r^5+4Mx^2-4My^2+2ry^2)\over{r^5(x^2-y^2)^2}} .}
Notice that the metric \schwarsdd\ is {\it not}
spherically symmetric,
in fact its only isometry is time translations. Neither is it
 asymptotically
flat. For large $r$, the $x$ dimension gets
squeezed and the other dimensions
behave like a $2+1$ dimensional space--time.
The surfaces $x = {\rm constant}$ are just
$2+1$ dimensional black holes away from the singularities
$\sin\theta=0$, $(y=\pm x)$ .
Again, it is straightforward to check that this solution
satisfies equations \einst--\dil\
thus providing new string vacua.

To find new solutions, we can certainly combine
both dualities above.  We can also consider different
coordinate systems. For instance, using the Eddington-Finkelstein
instead of the Schwarzschild coordinate system, we can
find its dual with respect to time translations.
It so happens that the dual metric
is identical to \schwarsdu, but now the new
solution has non--trivial torsion.  We have here come across
another general feature of duality symmetries: string
background solutions dual
to geometries related by a coordinate transformation are not
necessarily themselves related by a coordinate transformation .
They all however give the same path integral.

A similar analysis can be done for the 4D charged
dilatonic black holes of reference \gibbons. In this case the metric is
\eqn\cdbh{ds^2=-{(1-{2M/r})\over (1-Q^2/Mr)} dt^2 + {dr^2\over (1-{2M/r})
                   (1-Q^2/Mr)}
                    + r^2(d\theta^2 + \sin^2\theta d\phi^2)\ ,}
the dilaton field is $\Phi = -\log (1-Q^2/Mr)$ and the electric field
$F_{tr}= e^{\Phi}Q/(2r^2)$. It is very interesting
to note that the dual of this
solution with respect to time translations gives exactly the
same solution except that the mass parameter $M$ changes into
$ Q^2/2M$, therefore it relates
the black hole domain $Q^2<2M^2$ to
the naked singularity domain $Q^2>2M^2$.
In particular, the extremal solution $Q^2=2M$
is selfdual. We have then discovered
that 4D charged dilatonic black holes
share a similar property with
their 2D counterparts: a duality
transformation can give the
same geometry but interchange different
regions. The 4D solutions however
have more structure since another duality
transformation coming from the $SO(3)$
isometry can be performed. As in the Schwarzschild case,
this duality changes the angular part of the metric,
while leaving the $t,r$ components invariant:
\eqn\cdbhdd{\eqalign{
  ds^2=-{(1-{2M/r})\over (1-Q^2/Mr)} dt^2 + & {dr^2\over (1-{2M/r})
                   (1-Q^2/Mr)}\cr
   &\qquad\qquad + {1\over {r^2 (x^2 - y^2)}}\ [r^4 dy^2
                    + x^2 dx^2]\, \cr}}
with $\Phi' =  - \log [(1-Q^2/Mr)r^2 (x^2 - y^2)]$
and invariant electric field.
Again, the singularity structure
in the $r$ coordinate is the same as
for the original metric whereas there are new singularities at
$x=y$. Similar to the Schwarzschild dual, this solution is
not asymptotically flat and for large $r$ the geometry contracts
to a $2+1$ space.  The $x = {\rm constant}$ surfaces are again
$2+1$ dimensional black holes away from the singularities $y=\pm x$.

We should remark that all the backgrounds considered in this section
are solutions of the leading
order equations and then they are only
approximate solutions of the exact field equations, valid
in the domain of small curvature.  Nevertheless, exact solutions
with spherical symmetries surely exist for which the
formalism presented in this paper should apply.

\newsec{Conclusions}

We have shown that there is a new type of duality symmetry in
string theory associated with the existence of
a non--abelian group of
isometries on the target space--time in the
worldsheet $\s$--model, which reduces to
the usual target space duality
when the isometry group is abelian. We have
also shown that well known
properties of target space duality do not
extend to the general case.
In particular, the group of isometries ${\cal G}$
is not preserved in general under a duality transformation.
For instance,
the dual background of a model with
an abelian isometry also has an abelian isometry,
which can be enhanced
by a duality transfomation to a larger non--abelian
group of isometries.  This was the case of \dualUone\
which is dual to the Schwarzschild solution.
The non--abelian case is even more dramatic
since the original group of isometries
can completely dissapear after dualization.  Maybe
the most interesting open problem that this
work leaves is to find the mechanism by which
duality transformation can be performed
starting with backgrounds with no isometries
such as Calabi--Yau spaces
and gauged WZW models $G/H$ for which the
subgroup $H$ is nonabelian \refs{\gique,\bars}.

There are several other ways in which our work could be
extended.  In \vero, duality with respect to abelian
isometries was formally proved to relate two different
geometrical manifestations of a single conformal field
theory.  Probably, their analysis could be generalized
to the non--abelian case considered in this paper.
In particular, there are global considerations that must
be studied when identifying the path integral in \pathint\
with both dual theories.  For example, for a world sheet
with the topology of a torus, there are non trivial field
configurations which should be considered. In order for
the gauged theory to be equivalent to the original as
conformal field theories, it was shown in \vero\ that in
the abelian case, for compact isometries, the Lagrange
multiplier should have a specific periodicity.  In the
non--abelian case, the non trivial field configurations
ought to be considered too, though it is not known if
there exists a range for the Lagrange multipliers which
renders the theories equivalent.  All we can say at the
moment is that the dual geometries are not necessarily
equivalent as conformal field theories but that they are
related by an orbifold construction\footnote{$^*$}{We
thank E. Witten for stressing this point to us.}.

The duality due to an abelian $H$ is
known to map Bianchi identities of one
theory to field equations of the dual
and viceversa. From this a  set of transformations can be found
\refs{\gz,\cfg,\mv}
which {\it continuously} interpolates between Bianchi
identities and field
equations. This continuous symmetry, which is actually broken
non--perturbatively, is what generates the moduli space to which two
dual solutions belong.  It is also very powerful
when trying to find new solutions of
the string background equations \refs{\gmv,\sen}. It will
certainly be very interesting to investigate whether
these continuous transformations exist in the non--abelian case.
Also, at the moment we do not know what the full duality group is.
Of course, it has to include the group of
duality transformations with respect
to {\it all} possible subgroups of ${\cal G}$ and therefore
it must contain $SO(r,r,\IZ)$, where $r = {\rm rank} {\cal G}$.

Finally, one of our motivations to study
duality in 4D black holes, was
to analyze the effects of duality on the
singularities of those spaces.
At the moment we have not been able to find a map that
takes those singular regions to regular regions as
was the case with the 2D black holes \dvv\ and
3D black strings \refs{\hhs, \gique}.
Instead  there have appeared naked singularities. The  origin
of these new singularities
could be very interesting
to study in general. Also FRW cosmologies can be
analyzed in the present approach and dual geometries
to quasi realistic string cosmologies \tseyt\ could
be investigated. We hope to discuss some of these issues in a future
publication.



\bigskip
\bigskip
\bigskip
 {\bf Acknowledgements}

We would like to thank
L. Alvarez--Gaum\'e, C.P. Burgess, P. Candelas, J.--P. Derendinger,
E. Derrick, M. Duff, L.E. Ib\'a\~nez, K. Lee, D. L\"ust and P. Page
for very useful conversations. We especially
acknowledge P. Candelas for a careful reading of the manuscript and
the CERN Theory Division for hospitality.

\listrefs
\bye